\documentclass[runningheads]{llncs}

\usepackage{cite}
\usepackage{amsmath,amssymb,amsfonts}
\usepackage{algorithmic}
\usepackage{graphicx}
\usepackage[hidelinks]{hyperref}
\usepackage{textcomp}
\usepackage{wasysym}
\usepackage{lipsum}
\usepackage{tabulary}
\usepackage{multirow}
\graphicspath{ {images/} }

\newcommand*{\fullref}[1]{\hyperref[{#1}]{\autoref*{#1} \nameref*{#1}}}
\newcommand{\mypar}[1]{\vspace{1ex}\noindent\textbf{#1:}}

\begin{document}

\title{
    Revisiting Security Vulnerabilities in Commercial Password Managers%
\thanks{This is an accepted manuscript to appear in the proceedings of the 35th Int'l Conf.\ on ICT Systems Security \& Privacy Protection (IFIP SEC 2020), Maribor, Slovenia.}
}

\author{
    Michael Carr\inst{1} 
    \and 
    Siamak F.\ Shahandashti\inst{2}
    }

\institute{
    Piksel, York Science Park, YO10 5ZD, UK \\ 
    \email{mikey.carr@piksel.com}
    \and 
    Dept.\ of Computer Science, 
    University of York, YO10 5GH, UK \\ 
    \email{siamak.shahandashti@york.ac.uk}
    }

\maketitle


\begin{abstract}
    In this work we analyse five popular commercial password managers for security vulnerabilities. Our analysis is twofold. First, we compile a list of previously disclosed vulnerabilities through a comprehensive review of the academic and non-academic sources and test each password manager against all the previously disclosed vulnerabilities. We find a mixed picture of fixed and persisting vulnerabilities. Then we carry out systematic functionality tests on the considered password managers and find four new vulnerabilities. Notably, one of the new vulnerabilities we identified allows a malicious app to impersonate a legitimate app to two out of five widely-used password managers we tested and as a result steal the user's password for the targeted service. We implement a proof-of-concept attack to show the feasibility of this vulnerability in a real-life scenario. Finally, we report and reflect on our experience of responsible disclosure of the newly discovered vulnerabilities to the corresponding password manager vendors. 
\end{abstract}


\section{Introduction}
Passwords remain the dominant authentication mechanism in the digital realm despite their shortcomings. Furthermore, they are expected to persist as a primary authentication mechanism for the some time~\cite{herley2011research}. Among the tools that can greatly reduce the cognitive burden of remembering multiple passwords for multiple services are password managers. Hence, their use is strongly advocated by security experts, including the UK's National Cyber Security Centre~\cite{NCSC}. 


A password manager is an encrypted vault that stores any number of credentials for the user and is accessed by a single master password. In this context, a credential is a username-password pair that authenticates the user to a web-based service. Over and above individual use, a commercial password manager usually provides extra features, e.g.\ credential sharing and admin interfaces, and aims to increase enterprise security. Obviously, vulnerabilities in such an application provide opportunities for malicious actors to extract credentials, compromise commercial information, or violate employee privacy. Therefore, rigorous security analysis of password managers is crucial. 

Analyses focusing specifically on the security of password managers appear within academic literature and other less formal publications such as blogs as early as 2003. Each work usually reports one or more discovered vulnerabilities and how they could be exploited in an attack against certain password managers in the hope that password manager vendors eventually rectify these issues. However, it is not clear to what extent reported issues apply to other password managers and to what extent they are mitigated by corresponding vendors. In fact, there does not seem to be any reference that aggregates the major security vulnerabilities reported in the literature, and existing reports remain fragmented in multiple sources. In this work we attempt to address this gap as well as reporting new vulnerabilities we discovered in our analyses. 

We report the results of our work on analysing the security of the enterprise editions of five major password managers: LastPass, Dashlane, Keeper, 1Password, and RoboForm. These password managers were chosen after a rigorous selection process which considered popularity and features of the individual and commercial offerings of 19 password managers. Our contributions are threefold: 

1) We carried out a survey of both formally and informally published vulnerabilities of password managers, identified six main vulnerabilities, tested current versions of the five considered password managers against each vulnerability, and report our results whether each password manager is susceptible to each previously disclosed vulnerability. 

2) Through comprehensive systematic testing of mobile, desktop, and web applications (including browser extensions) of the considered password managers, we discovered four new issues that can lead to exploitable vulnerabilities, developed a proof-of-concept Android application to demonstrate how the most serious issue might be used in a real-world phishing attack, and report on whether each password manager is susceptible to each discovered vulnerability. 

3) Following the principle of responsible disclosure, we informed the corresponding password manager vendors of the newly discovered vulnerabilities. We report and reflect on our experience of interacting with these vendors. 

Many modern browsers provide password management services on the side. However, we focus on stand-alone password managers that provide a commercial offering for organisations and do not consider browser password managers. 

The rest of this paper is organised as follows: 
in Section~\ref{sec:litreview} we review the literature on password manager security; 
Section~\ref{sec:method} specifies our method for selecting and analysing the password managers considered; 
Section~\ref{sec:results} reports on our results on both previously disclosed and newly discovered vulnerabilities and discusses their feasibility and impact; 
Section~\ref{sec:disclosure} reports on our responsible disclosure to the corresponding vendors; and 
concluding remarks come in Section~\ref{sec:conclusion}. 

\section{Related Work}
\label{sec:litreview}
In recent years password managers have been analysed a multitude of times, both within and outside academia. Here we review the major reported vulnerabilities. 

\subsection{Autofill Vulnerabilities}
An area that has been of substantial interest to researchers is the autofill feature that password managers implement to increase their usability. A number of works have exploited poor implementation of autofill to extract user's credentials, in some cases automatically~\cite{silver2014password, blanchou2013password, 2013arXiv1309}. In~\cite{silver2014password}, a minimal survey implemented as an HTML form was
sent to users of multiple webmail services. The form contained a visible question along with invisible email and password input boxes. The idea was that the password managers would see the email as a login form with the webmail domain as origin and autofill the credentials for the webmail service. 
With auto-login enabled, merely opening the email would automatically fill in the credentials and submit the form in some webmail services, and in others the user was warned that a form is about to be submitted but would still be vulnerable if they clicked through the warning. 

\subsection{Web-Based Vulnerabilities}
A rigorous analysis of the security of five web-based password managers (including two we consider here, LastPass and RoboForm) was performed by Li et al.~\cite{emperors}. The authors found a diverse range of vulnerabilities ranging from classic web vulnerabilities such as cross-site scripting (XSS) and request forgery (CSRF) to more specific authorisation and user interface vulnerabilities. 
Notably the authors found that in some cases only authentication was carried out and not authorisation. This allowed an attacker registered with the password manager to successfully request a victim's password to be shared with another party. 


Bookmarklets were extensively used by password managers to provide browser integration when extensions (a.k.a.\ add-ons) were not available, e.g.\ in many mobile browsers. With the extended functionality of native APIs, and in view of the inherent vulnerability of bookmarklet code execution, password managers have moved on to providing separate applications on multiple operating systems, and either have already discontinued bookmarklet support or discourage its use. 

The vault encryption methods used in the enterprise versions of RoboForm and LastPass have been analysed in \cite{BCPM_Analysis} where the authors define a threat model that takes into account two forms of attackers with different capabilities: outsider attackers and insider attackers. 
The authors focus specifically on three forms of attack: brute force attacks, local decryption attacks, and request monitoring attacks. In LastPass, local decryption by outsider, brute force by outsider, and brute force by insider attacks were all capable of retrieving a user's master password. In RoboForm, vulnerabilities were found in local decoding by outsider, brute force by outsider, and server-side request monitoring by insider attacks. 

Dashlane was the subject of a security analysis in 2016~\cite{Dashlane_Analysis}. 
When attempting to log on, if an invalid username is entered, a message stating `Incorrect login' is shown, whereas if an incorrect password is entered, a message stating `Incorrect password' is shown. This indicates that a username is registered with Dashlane and would aid an attacker when attempting a brute force attack on usernames and passwords. Although an attacker would need access to the victim's devices as Dashlane uses two-factor authentication (2FA) for any new devices, a device authentication vulnerability meant that 2FA could be bypassed. 
This allows an unauthorised device to access passwords. 

\subsection{Non-Academic Sources}
In 2017, a vulnerability in the implementation of 2FA (via a QR code) in LastPass~\cite{Lastpass_2FA} was found. The issue was that the URL where the QR code was stored was a predictable hash of the user's master password. An attacker that is hoping to bypass 2FA will already know the victim's password and, therefore, is capable of accessing the QR code, which is needed to generate the valid temporary codes. 
The vulnerability was likened to a safe within a building that can be opened with the same key as that of the building door~\cite{Lastpass_2FA}. 
However, for the request to access the QR code to be valid, a user has to be authenticated, but it was shown that this could be defeated using a cross site request forgery vulnerability. 
LastPass have since patched this vulnerability. 

In 2016 a vulnerability was discovered in the LastPass extension URL parsing code used to decide on whether to autofill a website~\cite{MathiasKarlsson}. In short, it meant that a specially crafted URL was able to extract the credentials for arbitrary websites. Browsers treat a URL like \texttt{example.com/@twitter.com/@xyz.php} as from \texttt{example.com} while the extension treated it as from \texttt{twitter.com} since only the last occurrence of the \texttt{@} was considered. Hence, it was possible to confuse the extension and allow an attacker to identify the credentials for a targeted website. LastPass have since patched this vulnerability. 

There does not seem to be a survey of password manager security analyses bringing together an aggregated list of reported vulnerabilities. In this work, we address this issue. 
Note that there is a related line of research on the security of encrypted database formats used in password managers (see e.g.~\cite{gasti2012security}) which considers the threat model in which an adversary gets direct access to the password manager vault. We do not consider this threat model here. 

\section{Method}
\label{sec:method}
We specify the methods we used in each step of this work in the following. 


\subsection{Identification of Password Managers}
To ensure a wide range of password managers were considered, a comprehensive survey of individual and commercial password managers was undertaken. This took into account the number of users documented by the password manager vendors, install counts in application stores and recommendations by reputable websites such as PCMag.com. Further, less publicised products were identified by inputting terms such as `password manager' into web search engines. 

Once the search reached saturation, identified by a lack of new products appearing, each of the password managers listed was investigated and their features compared. To make our final selection, we considered two characteristics: popularity of the tool as an indicator of the number of users affected by a potential vulnerability, and richness of features, as an indicator of both desirability for companies as customers and at the same time diversity of attack vectors. 

\subsection{Identification of Previously Disclosed Vulnerabilities}
A comprehensive review of the literature on password managers was carried out. Besides, we examined general and tech news sources for any security issues reported for password managers. We limit our presentation to vulnerabilities that lend themselves to feasible and impactful attacks. 

To keep our survey focused on password manager specific vulnerabilities, we do not consider general web vulnerabilities such as XSS or CSRF for which there are standard recommended solutions. Furthermore, given the gradual phaseout and the inherent vulnerability of bookmarklets, we leave bookmarklet-specific vulnerabilities out of the presented results here. Ultimately, the results of our survey needs to be considered alongside the vulnerabilities listed by Li et al.~\cite{emperors}, and those considered by the literature on database format security (e.g.~\cite{gasti2012security}) for a more comprehensive view of password manager security. 

\subsection{Testing for Identified and New Vulnerabilities}
After selecting the products for testing, a two-week enterprise trial was started with each of the products consecutively. To begin, the systems were tested under normal operation to identify any abnormalities. This involved completing a large number of tasks using features that are available (to users as well as admins) in the enterprise editions of the software. 
A comprehensive list of all the operations that were performed is not presented here, but included the following: 
\begin{itemize}
\item logging in as a user and an administrator; 
\item adding a password through the vault and automatic capture features; 
\item sharing passwords between users; 
\item updating shared passwords and individual passwords on multiple devices; 
\item linking a personal account to the corporate account; 
\item analysing activity reports and their accuracy; and 
\item adding and removing users from groups and roles. 
\end{itemize}%

Following the initial testing of standard features, all of the password managers were tested against the identified previously disclosed vulnerabilities. Through checking previously disclosed vulnerabilities across all of the password managers in the sample, it is possible to establish whether patches have been correctly applied and whether the issues are common across password managers. 
The abnormalities discovered through the initial testing were then capitalised on through the development of proof of concept exploits. 

\mypar{Ethical Considerations}
Throughout this work, ethical considerations have been paramount. No live user account was used or attacked. All testing was carried out using accounts that belong to the investigators. All vulnerabilities have been responsibly disclosed to the vendors at least six months before publication. 

\section{Results}
\label{sec:results}
In this section we first give the detailed specifications of our test settings and then provide our results on testing the selected password managers against previously disclosed vulnerabilities and our discovery of new vulnerabilities. 


Our search for password managers identified 19 applications supporting most of the features that can be considered basic features for password managers, e.g.\ password capture and encrypted storage, password generation, mobile app, and autofill. 
Overall, we identified 27 features, including a number of desirable additional security features, e.g.\ two-factor and biometric authentication, and some that are especially desirable in a professional environment, e.g.\ password sharing, security breach alerts, admin console, and API provision. 
The full list of password managers and features we considered can be found in Appendix~\ref{sec:full-list}.

From the 19 password managers identified, those with the greatest popularity and richness of features were selected for testing. These password managers are LastPass, Dashlane, Keeper, 1Password, and RoboForm. 

The desktop components of the password managers were tested using a laptop running Windows 10 Enterprise version \texttt{10.0.14393} build \texttt{14393} and Chrome version \texttt{59.0.3071.115} where extensions were used. Any mobile components were tested using an Android 7.0 phone. Note that Windows, Android, and Chrome are respectively the current most widely-used OS, mobile OS, and browser worldwide. Tested password manager versions are shown in \autoref{tab:versions}. 


\begin{table}[ht]
\centering
\caption{Version numbers of the password managers tested.}
\label{tab:versions}
\begin{tabular}{@{\ }l@{\ } @{\ }c@{\ } @{\ }c@{\ } @{\ }c@{\ } @{\ }c@{\ } @{\ }c@{\ }}
\hline
\multirow{2}{*}{\textbf{Component}} & \multicolumn{5}{c}{\textbf{Password Manager}} \\
\cline{2-6} 
& \textbf{Dashlane} & \textbf{LastPass} & \textbf{Keeper} & \textbf{1Password} & \textbf{RoboForm} \\ 
\hline 
Chrome Extension & 4.8.2 & 4.1.60 & 10.8.1 & 6.6.439 & 8.3.7.7 \\ 
Android App & 4.17.0.1995 & 4.2.762 & 10.7.0 & 6.5.3 & 8.0.9 \\ 
\hline 
\end{tabular}
\end{table}


\subsection{Previously Disclosed Vulnerabilities}
\label{vulnerability summary}
Our survey resulted in the identification of six main issues that we list in this section. Our initial tests indicated that authorisation vulnerabilities discussed by Li et al.~\cite{emperors} appear to have been patched in all password managers we considered. Hence, we do not list those vulnerabilities here in the interest of conciseness. 

\newcommand{\tfas}{2FA Seed}
\mypar{Two-Factor Authentication Seed Vulnerability}
Disclosed for LastPass in which the seed for enabling 2FA was stored at a predictable URL~\cite{Lastpass_2FA}. This was tested on the password managers by initiating the 2FA set-up process and identifying whether the seed URL appear predictable based on any user information.

\newcommand{\elinss}{Element Inspect.}
\newcommand{\elinsl}{Element Inspection}
\mypar{Element Inspection Vulnerability}
Leakage of shared passwords through DOM element inspector tools such as Chrome's \texttt{Inspect Element}, as shown in the Dashlane security analysis~\cite{Dashlane_Analysis}. This can be tested by sharing limited access to a password with another user, which allows them to use the password but not see it. Following this, logging in as the other user and then using an element inspector on the shared password exposes the password. 

\newcommand{\regdiss}{Reg.\ Discovery}
\newcommand{\regdisl}{Registration Discovery}
\mypar{Registration Discovery Vulnerability}
An indication of whether a username is registered with the service through UI prompts, as shown in the Dashlane security analysis~\cite{Dashlane_Analysis}. This is tested through attempting to log in with an incorrect username and then incorrect password. 

\newcommand{\urlmm}{URL Mismatch}
\mypar{URL Mismatch Vulnerability}
Log in fields being filled with a username and password, despite the source and destination URLs not matching~\cite{blanchou2013password}. 

\newcommand{\httpsaf}{HTTP(S) Autofill}
\mypar{HTTP(S) Autofill Vulnerability}
Autofill policies do not distinguish between HTTP and HTTPS when attempting to fill a credential that has been stored with HTTPS on an HTTP version of the site~\cite{blanchou2013password}. This would enable a man-in-the-middle attacker to impersonate an HTTP version of a popular website and steal user credentials originally stored for the HTTPS version. 

\newcommand{\igsubds}{Ignoring Subdom.}
\newcommand{\igsubdl}{Ignoring Subdomains}
\mypar{Ignoring Subdomains Vulnerability}
Subdomains are ignored when filling passwords~\cite{blanchou2013password}. This is tested with the university websites: \texttt{york.ac.uk} and \texttt{cs.york.ac.uk}. An attacker in a subdomain can hence steal user credentials for the parent domain or other subdomains. This is an issue in many websites such as forums and blogs where different subdomains host different services. 

\mypar{Summary}
\autoref{tab:sum} shows the results of testing the five password managers against the vulnerabilities listed above. These vulnerabilities were tested using the same processes and resources for all the password managers. As can be seen from the table, at least one of the password managers is vulnerable to every single issue apart from \tfas\ vulnerability. The tested password managers are most vulnerable to \urlmm, \httpsaf, and \igsubdl\ vulnerabilities, with all but one of the managers being susceptible to \urlmm\ and all to \httpsaf\ and \igsubdl\ vulnerabilities. All of these vulnerabilities concern the web interfaces, and more specifically the autofill feature, which has been an area of focus for previous works. Hence, it was hoped that vendors had responded by making their software resilient to such attacks. However, this appears not to be always the case. 

\begin{table}[th]
\caption{Previously disclosed vulnerabilities analysed against the password managers tested. A $\CIRCLE$ indicates the application is vulnerable, a $\Circle$ indicates it is not.}
\label{tab:sum}
\centering
\begin{tabular}{@{\ }l@{\ } @{\ }c@{\ } @{\ }c@{\ } @{\ }c@{\ } @{\ }c@{\ } @{\ }c@{\ }}
\hline
\multirow{2}{*}{\textbf{Vulnerability}} & \multicolumn{5}{c}{\textbf{Password Manager}} \\
\cline{2-6} 
& \textbf{Dashlane} & \textbf{LastPass} & \textbf{Keeper} & \textbf{1Password} & \textbf{RoboForm} \\ 
\hline 
\tfas & $\Circle$ & $\Circle$ & $\Circle$ & $\Circle$ & $\Circle$ \\ 
\elinsl & $\CIRCLE$ & $\CIRCLE$ & $\CIRCLE$ & $\Circle$ & $\Circle$ \\ 
\regdisl & $\CIRCLE$ & $\Circle$ & $\Circle$ & $\Circle$ & $\Circle$ \\  
\urlmm & $\CIRCLE$ & $\Circle$ & $\CIRCLE$ & $\CIRCLE$ &  $\CIRCLE$\\ 
\httpsaf & $\CIRCLE$ & $\CIRCLE$ & $\CIRCLE$ & $\CIRCLE$ & $\CIRCLE$ \\ 
\igsubdl & $\CIRCLE$ & $\CIRCLE$ & $\CIRCLE$ & $\CIRCLE$ & $\CIRCLE$ \\ 
\hline 
\end{tabular}
\end{table}

\subsection{Discovered Vulnerabilities}
Our feature testing flagged issues for further investigation. Here we present a developed proof-of-concept attack and three other vulnerabilities of the tested password managers. Unlike previous vulnerabilities, the ones we discuss here do not only concern web interfaces and some are related to mobile apps. 

\mypar{Phishing Attack}
Both the 1Password and LastPass Android applications were found vulnerable to a phishing attack. The issue discovered was that both applications use weak matching criteria for identifying which stored credentials to suggest for autofill. This allowed for a rogue application to impersonate a legitimate one simply by crafting the package name to be identical. A developed proof-of-concept attack is described in detail below for LastPass but essentially the same attack applies to 1Password. 

To identify the process used when matching an application and credentials, a blank login screen was created. After selecting the \texttt{Add Login} option in the LastPass pop-up, the URL shown in the LastPass application is the package name of the application developed. This indicated that the matching criteria employed by LastPass is based on the package name of the application only.

\begin{figure}[t]
\centering
\begin{minipage}[t]{0.40\linewidth}
    \centering
    \includegraphics[width=0.75\textwidth]{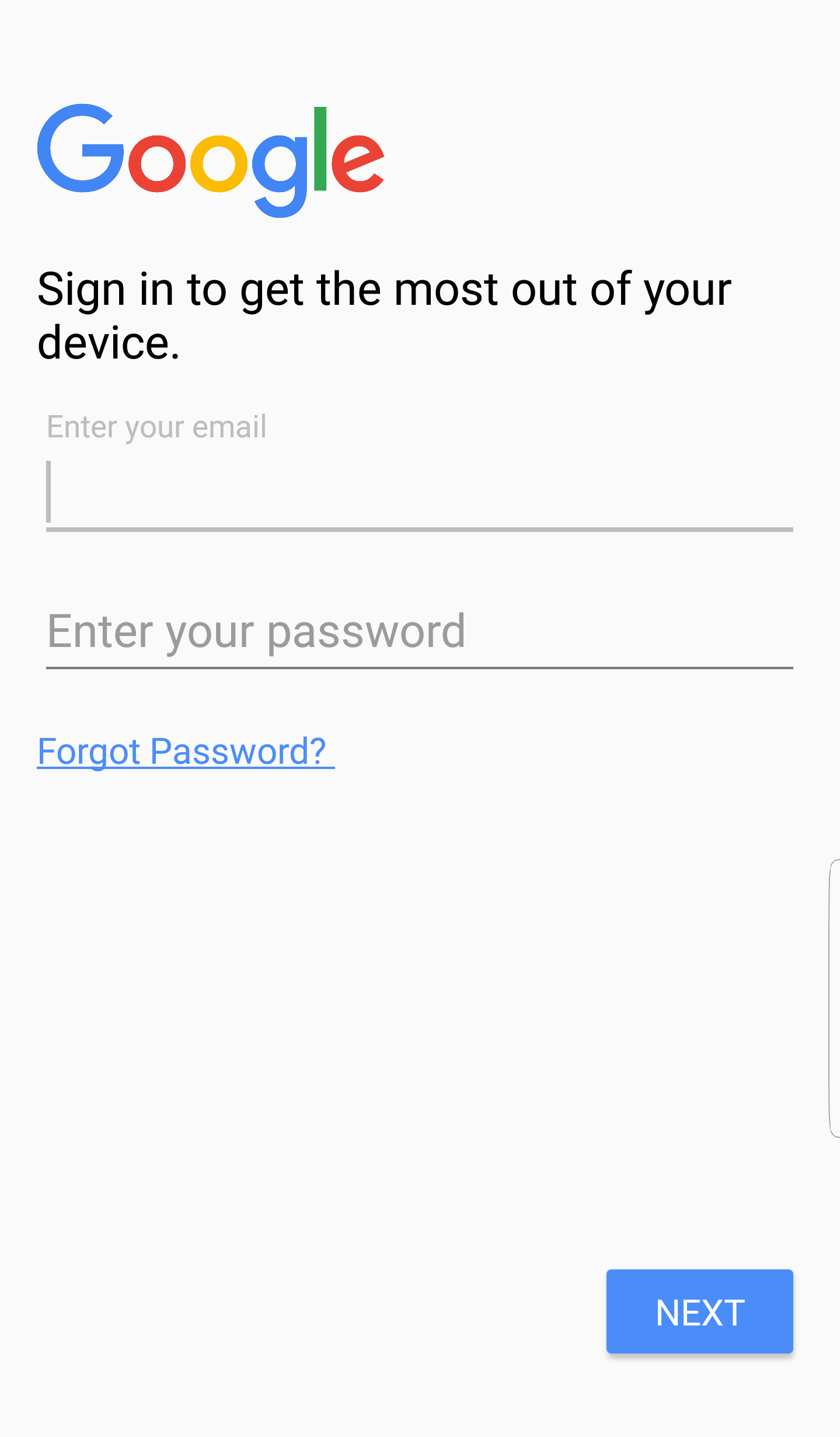}
    \caption{Login screen of the malicious app mimics that of Google}\label{fig:googleLogin}
\end{minipage}
\hspace{0.2cm}
\begin{minipage}[t]{0.40\linewidth} 
    \centering
    \includegraphics[width=0.75\textwidth]{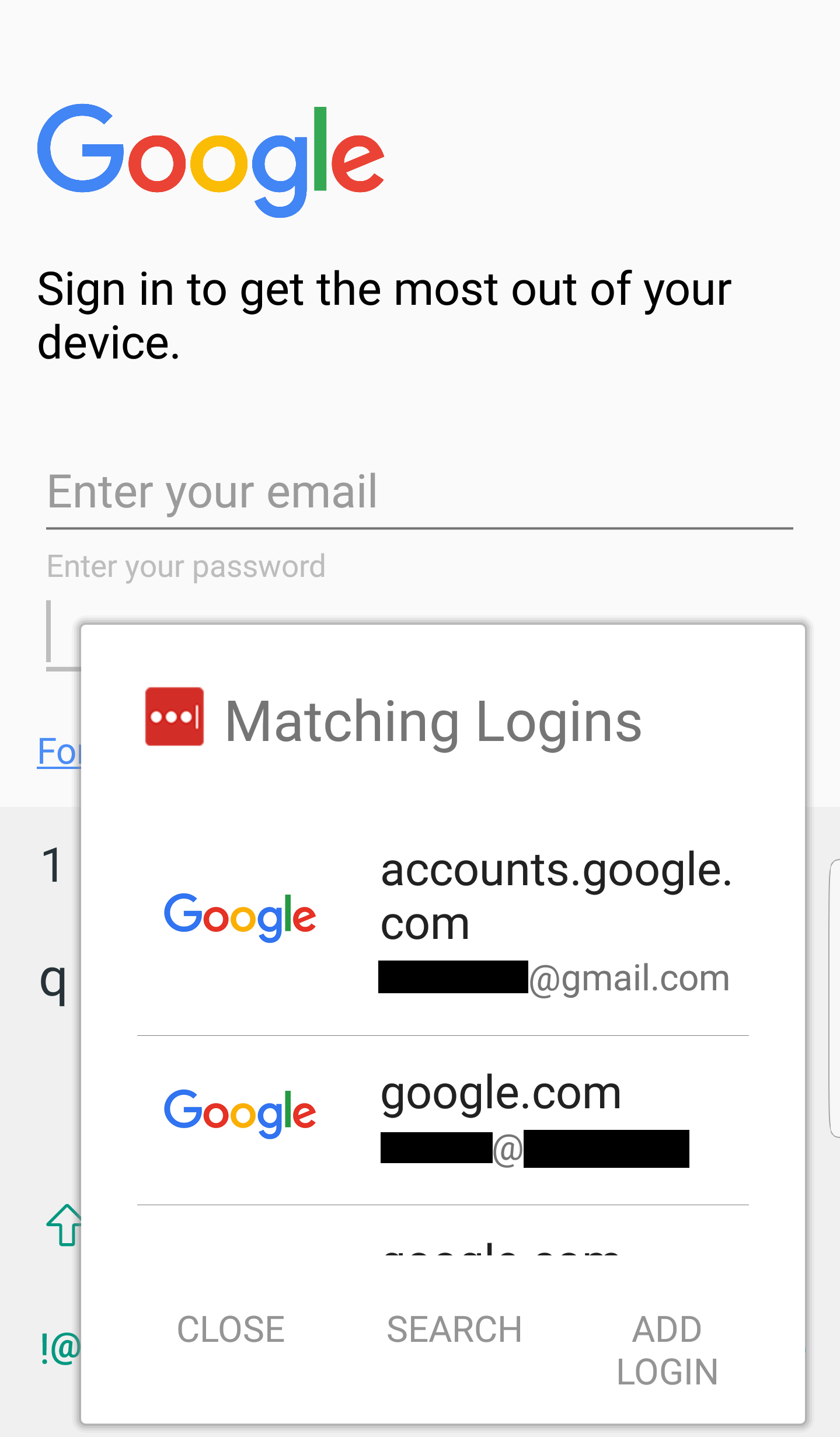}
    \caption{LastPass showing Google credentials for the malicious app} \label{fig:popup}
\end{minipage}        
\end{figure}

After discovering how LastPass matches applications and credentials, a malicious app was developed with the package name of \texttt{com.google}. This app had a login screen, shown in \autoref{fig:googleLogin}, that was designed to mimic that of the official Google login screen and thereby be hard to distinguish. The weak matching employed by LastPass means that when the malicious app is launched, LastPass will offer to autofill the login page with Google credentials stored in a user's vault. This can be seen in \autoref{fig:popup}. In our proof-of-concept attack, after a victim selects their credentials from the LastPass pop-up and taps the \texttt{Next} button, the credentials are sent across to a server and stored. Hence, as long as the victim is tricked into installing and launching a malicious application, their credential can be stolen easily leveraging the weak matching used by 1Password and LastPass. 

The attack developed here succeeds if the following conditions are met. 
Firstly, the malicious app needs to be installed on the victim's device. Attackers might achieve this by either getting around app store security mechanisms (see e.g.~\cite{oberheide2012dissecting} in the case of Google Play Store) or otherwise fooling the victim into sideloading the app onto their device. 
This could be done in combination with another phishing attack, for example, sending an email stating the targeted service's application requires an upgrade. 
Secondly, the victim needs to be a user of the vulnerable password managers and using the LastPass or 1Password autofill prompt, although other users may be fooled and enter their password manually. 
Finally, the user needs to have credentials for the target application, in this case, Google, in their vault. 
Having said these, if an organisation is identified as a commercial user of a vulnerable password manager along with other services (e.g.\ Google email service), the latter two conditions are met and a large-scale phishing campaign may be launched against the organisation employees which one expects to have some degree of success in stealing employee credentials and thereby potentially compromising organisation security. 
The suggested mitigation for this vulnerability is for password managers to apply more strict matching criteria that is not merely based on an app's purported package name. 

\mypar{Clipboard Vulnerability}
A crucial usability feature of password managers is the ability to autofill credentials on a website. 
While autofill performs as expected on an overwhelming majority of websites across all the password managers tested, occasionally it would not. When the autofill feature does not work, password managers often provide the option to copy credentials to the clipboard.

It was discovered during the initial testing phase that the tested password managers do not provide enough protection surrounding copying sensitive items to the clipboard, except 1Password. 
Standard computer security advice recommends that a user locks their machine as soon as they leave it unattended and if a user was to follow this advice, the risk associated with leaving passwords in the clipboard should be reduced in theory. However, Windows 10 allows access to the clipboard of a locked machine~\cite{clipboard_lock_screen}. This allows pasting in the value of the clipboard in cleartext by an adversary that may be a person with physical access to the machine or an application running on the machine. Although the attack will not be aware as to what account this password is associated with, they can try the credentials with a precompiled list of websites for which autofill is known not to work. 
The suggested mitigation for this issue would be for the password managers to provide an option to clear the clipboard after a set amount of time. 

\mypar{PIN Brute Force Vulnerability}
To ease authentication to the Android applications, some password managers allow for a user to set a four digit PIN to access the application. This removes the need for a user to enter a long, complex master password every time they wish to enter their vault. 

It was discovered during testing, that the RoboForm and Dashlane Android applications do not correctly implement a persistent counter on the number of times an incorrect PIN can be entered when trying to access the application. It is possible to attempt two PINs consecutively, remove the application from the recent application drawer, then try a further two PINs. Both PINs were four digits long and therefore, have 10,000 combinations. Through extrapolation of manual testing, it is estimated that even a manual random guessing attack is on average expected to find a randomly selected PIN in 2.5 hours. If the attacker was to factor in common PINs the results in~\cite{bonneau2012birthday} suggest that the attack time would reduce to approximately 1.5 hours, and if the birth date of the victim is known, around 8\% of the PINs are expected to be found within the first six guesses. We did not fully automate this attack, but we expect an automated attack to take considerably less time to brute force the PIN. 


This attack has the potential to be catastrophic for the victim. A malicious attacker would have full access to the application, providing there is no prompt for the user to re-authenticate using something other than the PIN. Access to the application in both Dashlane and RoboForm enables the user to view, modify, or delete records within the password manager's vault. 

The suggested mitigation for this issue is to implement a persistent counter that is not reset when the application is removed from recent applications. 

\mypar{Possible Brute Force via Extension}
All password managers tested provide browser extensions that enable access to their vaults. To log in after the extension has initially been locked, the user only needs to enter their master password. 

Our tests suggested that Keeper, Dashlane and 1Password might be vulnerable to a UI driven brute force attack when entering the master password. This was because we did not observe any measures in place to halt the authentication process for 10 tested unsuccessful login attempts. This suggested that it might be possible to mount a dictionary attack against a user's account. 



Technically, the attack should be identified and halted by the password manager vendors. When testing, ten incorrect passwords were attempted against accounts with each of the password managers with no indication that a count on the number of incorrect attempts was being kept in any of the five. RoboForm implements a five second time delay after three incorrect passwords and in LastPass, there are multiple clicks required between entering successive passwords. These measures slow down a possible attack, but do not prevent it. 

Due to ethical reasons, the number of passwords attempted was considerably less than would be required in a dictionary or brute force attack, and it is possible that the vendors implement measures to identify larger number of login attempts. Hence, we regard this only as an indicator of a possible vulnerability requiring further investigation. 
The standard mitigation would be to lock the a user's account following a number of incorrect login attempts. 


\mypar{Summary}
\autoref{tab:dev_vulns} shows a summary of the susceptibility of five tested password managers to the discovered vulnerabilities. The reported issues are categorised as an (implemented) attack, two vulnerabilities, and a potential vulnerability. 


\begin{table}[th]
\centering
\caption{New vulnerabilities discovered and their contexts: mobile app (app) or browser extension (ext). A $\CIRCLE$ indicates no countermeasures observed (hence application is or could be vulnerable), a $\LEFTcircle$ indicates only partial countermeasures observed, a $\Circle$ indicates sufficient countermeasures observed (hence application is not vulnerable).}
\label{tab:dev_vulns}
\begin{tabular}{@{\ }l@{\!\!\!} @{\ }c@{\ } @{\ }c@{\ } @{\ }c@{\ } @{\ }c@{\ } @{\ }c@{\ }}
\hline 
& \multicolumn{5}{c}{\textbf{Password Manager}} \\
\cline{2-6} 
& \textbf{Dashlane} & \textbf{LastPass} & \textbf{Keeper} & \textbf{1Password} & \textbf{RoboForm} \\ 
\hline 
\textbf{Attack:} & & & & & \\ 
Phishing (app)    & $\Circle$ & $\CIRCLE$   & $\Circle$ & $\CIRCLE$   & $\Circle$ \\ 
\hline 
\textbf{Vulnerabilities:} & & & & & \\ 
Clipboard (ext)   & $\CIRCLE$ & $\CIRCLE$ & $\CIRCLE$   & $\Circle$   & $\CIRCLE$ \\ 
PIN Brute Force (app)    & $\CIRCLE$   & $\Circle$ & $\Circle$   & $\Circle$   & $\CIRCLE$ \\ 
\hline 
\textbf{Potential vulnerability:} & & & & & \\ 
Pwd Brute Force (ext)     & $\CIRCLE$ & $\LEFTcircle$ & $\CIRCLE$ & $\CIRCLE$ & $\LEFTcircle$ \\
\hline   

\end{tabular}
\end{table}

\mypar{Discussion on Feasibility and Impact}
\label{sec:discussion}
To further contextualise the discovered attacks and vulnerabilities here, we discuss their feasibility and potential impact. In our analysis, we adopt an approach similar to that of the Common Vulnerability Scoring System (CVSS) industry standard (see \url{www.first.org/cvss}) to make it easier to translate our discussions here to CVSS scores if necessary. 

The phishing attack discovered and developed here is a highly feasible attack. It does not require any physical or privileged access to the device, but it does require user interaction. As long as a victim is tricked into installing a malicious app it will be able to present itself as a legitimate and rather indistinguishable option on the autofill prompt and have a high chance of success. The impact of the attack is limited to the loss of a single credential at a time, although this may be a highly valuable credential (such as an email password) that could possibly enable access to further accounts. The loss of a single credential enables the attacker to get access to the compromised account, change content, and make content unavailable by changing the credential, so the account is potentially affected on all three aspects of confidentiality, integrity, and availability. 

The clipboard vulnerability is limited to an attacker with physical access to the device, however it does not require any privilege. This is an opportunistic attack that requires a specific uncommon user action, i.e.\ copying credentials to the clipboard. The impact is loss of the credential, hence affecting confidentiality, integrity, and availability for the compromised account. 

The PIN brute force vulnerability may also enable an opportunistic attack. It requires physical access to the victim's device. Although it does not require any further privilege of user interaction. The impact of the attack is much more severe than the previous two attacks since a successful attacker gets access to the entire password manager vault and any services whose passwords are managed by the password manager. This means that a successful attacker may freely access and modify the contents and credentials of all the managed accounts and hence it amounts to severe possible loss of confidentiality, integrity, and availability. 

Finally, the possible brute force via extension vulnerability would only require network access to the victim's device and no specific privilege or user interaction. However, since a password is being brute forced, the probability of success is typically less than the case where a PIN is targeted. Nevertheless, the low complexity and remote executability of this attack make it a highly feasible attack which if not mitigated can be exploited rather comfortably. The attack also has the potential of a severe impact in the form of loss of the master password that enables the attacker to access all the accounts managed by the password manager. Hence, confidentiality, integrity, and availability may be all severely impacted. Perhaps the only limitation of this attack is that it can be, at least in theory, easily identified through detecting higher than usual frequency of attempts to gain access to the targeted password manager account(s). 

\section{Responsible Disclosure}
\label{sec:disclosure}
In this section we go through our disclosure of the vulnerabilities to the vendors and their responses. 
We started discovering the vulnerabilities discussed here in 2017. After confirming the persistence of issues and developing and successfully testing our proof-of-concept attack, we started notifying the five vendors in 2018. For more severe vulnerabilities, we emailed the technical team and in some cases were asked to follow up the conversations through dedicated vulnerability reporting programs. For less severe issues, vendors were contacted through support tickets on their website. We continued our disclosure until late 2018 when we notified the vendors of our intention to publish our results in the form of an academic paper after a further public non-disclosure period of six months. 

In general we found all the five vendors quite responsive. However, only a few disclosures resulted in a fix to be rolled out. This was due to many of the disclosed issues being classified as low priority. In the following we discuss some of the more notable interactions we had with the vendors. 

The phishing vulnerability was disclosed to LastPass via their vulnerability reporting system and to 1Password via email. At LastPass, it was marked as `External Behaviour $>$ Browser Feature $>$ Autocomplete Enabled' which has a priority of 5 (lowest priority) according the Bugcrowd Vulnerability Rating Taxonomy (\url{https://bugcrowd.com/vulnerability-rating-taxonomy}) and therefore was assigned a response of `Won't Fix'. LastPass had no further comment. 

The clipboard vulnerability was communicated to all vendors affected. Dashlane stated that unlike mobile OSs such as Android, Windows does not provide any expiration mechanism for partial removal of data on clipboard and hence the only way to remove data from clipboard would be to delete all data on it. This would make such a solution quite invasive. Keeper responded that clipboard expiration was supported on iOS devices only. RoboForm told us that ``if you are using copy/paste actions for inserting some passwords from the RoboForm Editor, you will need to clear the clipboard manually.'' 

The PIN brute force vulnerability was disclosed to the affected vendors including 1Password via their  reporting system. 1Password fixed the issue within 11 days of reporting and emphasised that this vulnerability requires access to an unlocked Android device to be exploited. Dashlane told us they will add a persistent counter and later added that ``this issue requires access to an unlocked android device to be exploited'', implying a low priority. 

The extension brute force vulnerability was disclosed to all vendors. 1Password responded that it would be infeasible to guess a user's master credentials. Keeper indicated that a mechanism is in place to lock accounts after a threshold number of (20, and subsequent to our testing 10) unsuccessful login attempts. 

\section{Conclusions}
\label{sec:conclusion}

This work has analysed and reported on vulnerabilities in commercial password managers through two distinct avenues: testing previously disclosed vulnerabilities and developing exploits for newly discovered vulnerabilities. Many of the previously reported vulnerabilities have been found to persist in popular password managers. Furthermore, four new vulnerabilities were found through extensive testing and responsibly disclosed to the corresponding vendors. Some were fixed immediately while others were deemed low priority. 

In our correspondence with password manager vendors we saw both positive and negative sides to how they deal with vulnerability disclosure. On the positive side, vendors appear to be quite responsive and issues deemed high priority or easily rectifiable are fixed promptly. On the negative side, issues assessed as low priority appear to be considered non-issues and rather too easily dismissed. 

We acknowledge that some issues, e.g.\ the clipboard vulnerability, do not have an easy fix and vendors faced with a choice between leaving a low priority issue and applying a fix that has side effects may choose the former. 

A possible future direction would be developing rigorous security models and canonical security tests for password managers. The newly discovered vulnerabilities is this work were all user interface related vulnerabilities, as in they were discovered by testing the applications under typical operation scenarios. Such functionality tests along with further analyses focusing on architecture and processes may serve as a basis for standard tests. 

\bibliographystyle{splncs04}
\bibliography{references}

\appendix

\section{Full List of Password Managers and Features}
\label{sec:full-list}
The 19 password managers we considered are (alphabetically): 
1Password, Dashlane, EnPass, KeePass, Keeper, LastPass, LogMeOnce, mSecure, Password Boss, Password Manager Pro, Password Safe, PasswordState, RoboForm, SplashID, Sticky Password, TeamPassword Manager, TeamsID, TrueKey, and Zoho Vault. 

The 27 features considered are as follows (in no specific order): 
Mobile Applications, Extension Support, Bookmarklet, AutoFill, Form Fill, Password Capture (automatically saving entered passwords), Password Generation (generating new strong passwords), Multi Device Sync, Biometric Authentication, AES-256 Encryption, Vault Backup/Export Functionality, Password Sharing (with other users), Two Factor Authentication, Store data types other than passwords, Automatic Password Changes (e.g.\ on a regular basis), Portable App, Security Breach Alerts, Password Change Listener, Commercial Offering (for organisations), Admin Console (for organisation admins), Active Directory Integration, Reports and Auditing, Personal and Business Password segmentation (on the same app), Group Password Sharing, Manage User Account and Roles (for organisation admins), Custom Password Policies, and API Provision. 

\end{document}